# Evidence suggests that SARS-CoV-2 rapid antigen tests provide benefits for epidemic control - observations from Austrian schools

Jitka Polechová[1], Kory D. Johnson[2], Pavel Payne[3], Alex Crozier[4], Mathias Beiglböck[1], Pavel Plevka[3], Eva Schernhammer[5]

[1]Department of Mathematics, University of Vienna, Austria
[2]Institute of Statistics and Mathematical Methods in Economics, TU Wien, Austria
[3]CEITEC, Masaryk University, Brno, Czech Republic
[4]Division of Biosciences, University College London, London, UK
[5]Division of Epidemiology, Center for Public Health, Medical University of Vienna, Austria

**Corresponding authors:**
Jitka Polechová, jitka@univie.ac.at
Kory D. Johnson, kory.johnson@tuwien.ac.at

**Abstract:**
Rapid antigen tests detect proteins at the surface of virus particles, identifying the disease during its infectious phase. In contrast, PCR tests detect viral genomes; they can thus diagnose COVID-19 before the infectious phase but also react to remnants of the virus genome, even weeks after live virus ceases to be detectable in the respiratory tract. Furthermore, the logistics for administering the tests are different, with rapid antigen tests being much easier to administer at-scale. In this article, we discuss the relative advantages of the different testing procedures and summarise evidence that shows that using antigen tests 2-3 times per week could become a powerful tool to suppress the COVID-19 pandemic. We also discuss the results of recent large-scale rapid antigen testing in Austrian schools. While our report on testing predates Delta, we have updated the review with recent data on viral loads in breakthrough infections and more information about testing efficacy, especially in children.

**Keywords:**
COVID-19, SARS-CoV-2, lateral flow device, public health, repeat testing, large-scale testing

Rapid SARS-CoV-2 antigen tests are now widely available and have been provided free of charge for home-testing in Austria since March 1, 2021. We focus specifically on their comparison to PCR tests, which are the reference standard for diagnosing COVID-19. Frequent testing can improve pandemic control by lowering the transmission rate and the effective reproduction number. This potential can only be fully realized if these tests are used correctly and if the public fully understands both their capabilities and their limitations. A particularly important factor is that even when used in a supervised manner, rapid antigen tests are less sensitive than PCR tests.

In general, a virus can be detected by looking for its genetic material (DNA or RNA) or by detecting viral antigens which are present at the surface of the virus. As opposed to antibody tests, which detect antibodies from previous infection, antigen tests are used to detect people who are currently infected – and infectious. A few dozen to a hundred virus particles are sufficient for a rapid antigen test to detect SARS-CoV-2 [1]. Antigen tests are now readily available to the public and can provide results within 15 minutes. Therefore, they are a useful tool for rapidly identifying and isolating positive, infectious cases in order to reduce further transmissions. In contrast, a PCR test can detect the virus at even lower concentrations, as the PCR cycler multiplies fragments of the viral genetic material, but the execution of a PCR test, from sample collection to delivery of the results, is time consuming. Epidemic control relies on testing being done affordably at-scale, so that a significant proportion of the population can test multiple times per week.

In practice, two factors of PCR testing enable further transmissions: first, there is often a significant lag between PCR-test and result; second, the test result is used even later, e.g., admission to an event one to two days after the sample was taken. Therefore, one must consider test *effective* sensitivity *at the time of use*, not merely at the time of testing [2].

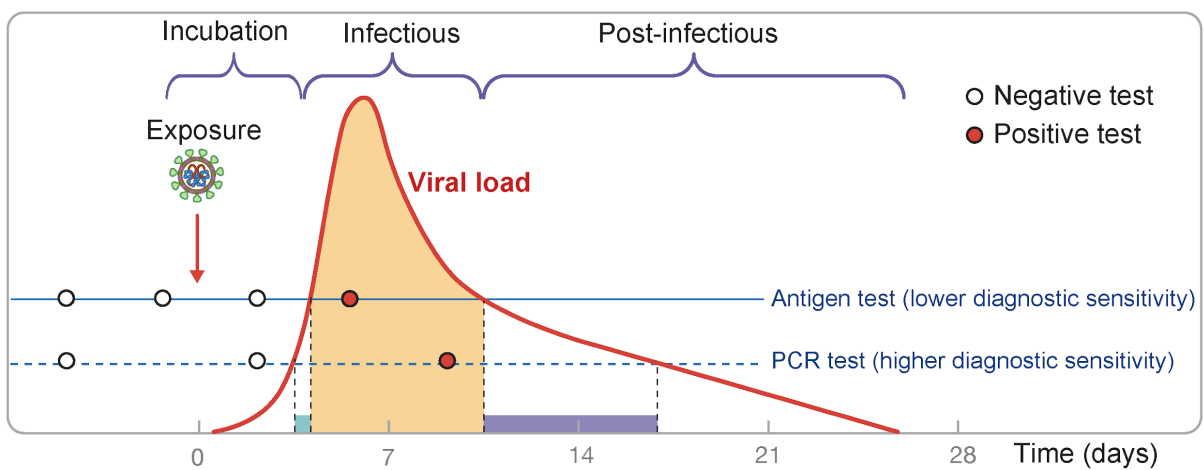

**Figure 1**. Despite having lower diagnostic sensitivity, rapid antigen tests identify the vast majority of infectious cases. Frequent testing is essential to identify these cases early and to efficiently limit the spread of the virus in the population. Figure, first published in [3], is modified from [4] with permissions; viral load is shown on a logarithmic scale.

Within the first two to three days after infection with SARS-CoV-2, neither PCR nor rapid antigen tests can detect the virus (Figure 1), as its concentration is too low [5]. During the subsequent steep rise in virus concentration, a PCR test detects infection a little earlier than a rapid antigen test [6]. This period, however, is short, lasting around a day or even less [2,7]. It is primarily within this 24-hour period that PCR tests provide additional benefit compared to antigen tests for the purpose of reducing further transmissions. While detecting infections early is key, the benefit of a PCR test to detect cases at a lower viral load (earlier in the infection) is often negated by the significantly longer turnaround time of the PCR-test result. While it is possible for an express PCR test to be returned in a few hours (such as at hospitals or airports), in general practice, the lag between test and result tends to be about a day or even longer.

As viral load and infectiousness increase, the ability of the antigen test to detect the virus also increases. The proliferation phase (day 0 to 7), where the viral load increases, does not seem to substantially differ neither between asymptomatic and symptomatic individuals nor between vaccinated and unvaccinated individuals. The peak viral concentration is reached in about 7 days after the infection [2,8–10]. The viral load declines slower, on average, in symptomatic individuals [2], with more severe infections taking significantly longer to clear [8]. Conversely, in vaccinated individuals, the clearance is faster than in unvaccinated individuals, regardless of the presence of symptoms. The faster clearance time of asymptomatic or vaccinated individuals [2,9,11] may explain the reported lower peak $C_t$ values in studies with less dense sampling of the asymptomatic individuals [12,13].

The "infectious period" is shown in orange in Figure 1 and usually lasts 5 to 8 days [8–10]. The majority of new infections occur before or soon after the peak viral load is reached [14,15]. We refer to a person as "infectious" when a virus sampled from the respiratory tract can be propagated in a cell culture. Recent data indicate that in vaccinated individuals, the presence of viable virus declines even faster (for the same viral load approximated by $C_t$ value) [16], and the infectious period is thus likely to be shorter. Infectiousness and viral load, while correlated, are not synonymous. The short period of high infectivity explains why frequent testing is essential for suppressing the spread of COVID-19: testing less frequently than once per week has little effect on new infections as most cases will be detected too late.

In the last phase of disease progression (shown in purple), virus can no longer be propagated [8,17]. Without any live virus, a person is no longer infectious and the antigen test will likely be negative [17]. In contrast, PCR tests may still produce positive results in this case, because viral RNA remains in the respiratory tract at detectable levels.

The diagnostic sensitivity of a test is defined as its ability to detect infection by a virus. A PCR test remains the reference standard for diagnosing a COVID-19 infection. Its use as the 'gold' reference standard is neither the result of nor implies that it has a 0% false negative rate [5]. It relies on the viral RNA (with the PCR targets) being present in the tissue from which it was sampled, the quality of the sampling process, and no major error in further processing [18–20]. Secondly, $C_t$ values used as a cut-off to declare an active infection (such as $C_t$ < 30) lead to a different probability of an active infection in the proliferation vs. clearance phase [2], and are not in general equivalent between labs.

Importantly, the infectious phase (orange in Figure 1) is considerably shorter than the overall time during which a PCR test can detect viral RNA in an infected person. To suppress an epidemic, the ability to detect and isolate an infectious person before they infect others is decisive: timing is paramount. Identifying a case in the post-infectious phase is important for diagnosis but not from a public health perspective, i.e., for preventing secondary infections.

Thus, the higher diagnostic sensitivity of PCR tests does not directly translate to improved pandemic control: [7,17] demonstrate that the *frequency* of testing is more important than the modality of testing. For epidemic control, we are interested in an effective sensitivity [2] from the time point the test was taken, to some time point in the future when the result is delivered and used. Smith et al. [17] indicate that antigen tests and PCR tests have similar power to identify "individuals before or during the period when infectious virus was detectable in nasal samples" when the interval between tests is one day shorter for antigen tests than for PCR tests (Fig 3B). In addition, they showed that when testing at least every 3 days, both PCR tests (nasal and saliva) and rapid antigen tests (Quidel SARS Sofia FIA) find over 98% of infected cases during the course of the infection. While antigen tests are effective in detecting an active infection, nasal PCR tests perform significantly better in the pre-infectious phase: the PCR test has a reported sensitivity of 70%, compared to 40% for an antigen test, to detect an infection within two days before virus could be cultivated from a nasal swab [17].

It is therefore important to prioritize systems which allow for more frequent testing or testing immediately before an event with high transmission risk. Larremore et al. [7] indicate that if half of the population would self-test every 3 days with a rapid antigen test and (immediately) isolate in the case of a positive result, we could achieve approximately a 40% reduction of the effective reproduction number R. For example, an R of 1.3 could be reduced to 0.8, and the epidemic would dissipate. Furthermore, it has been reported that over 80% of new infections are caused by fewer than 20% of cases [21–23]. Such so-called 'superspreading' is caused by 'superspreaders' who have both a large number of contacts and often a higher viral load (at the time of superspreading). These cases are more easily identified using rapid antigen tests as they would typically have a substantial viral load - and identifying them early would yield a large reduction in further transmissions. As such, there is an additional benefit in testing people with many contacts even more frequently.

With proper instruction, the effect of sampling (nasopharyngeal vs. nasal/throat swabs) on the detection of infectious individuals can be minimised. A German study found that out of 30 individuals with high viral loads (more than 10 million of viral RNA per swab; $C_t < 25$) all were correctly identified through rapid antigen tests with professionally administered nasopharyngeal swab, and 29 were identified through rapid antigen tests with self-performed anterior nasal swab [24]. The authors also concluded that "supervised self-sampling from the anterior nose is a reliable alternative to professional nasopharyngeal sampling using a WHO-listed SARS-CoV-2 [rapid antigen test]".

In order to interpret the outcome of wide scale antigen testing, it is helpful to have a rough estimate for how many positive cases one expects to find. We consider testing a randomly selected person from the general population whose infection status is unknown but who is currently not in quarantine, i.e., a person who does not suspect to be infected at the time of testing. Bearing in mind that we only aim to understand the order of magnitude of antigen-detectable cases, multiple lines of reasoning suggest that it is very low. First, a detailed epidemiological model fitted to Austrian data on February 15, 2021 estimates that 0.09% of the population is infected but not in quarantine [25]. Second, a back-of-the-envelope calculation arrives at a similar conclusion. Assume that there are around 1,500 new cases reported per day (as observed for much of February in Austria, which was before the arrival of the Delta variant) and that there is a case detection rate of 50% [26]. In this scenario, 1,500 people become infectious every day and do not quarantine.

Assuming that they are infectious for a week (in line with Figure 1), this results in around 10,000 undetected infectious individuals in Austria – i.e., cases which can be detected via an antigen test. Since the population of Austria is 8.9 million, the probability that a person in Austria tests positive via an antigen test is therefore approximately 0.1% under the given scenario. (Note that this calculation has been simplified by ignoring errors that are hard to quantify and have countervailing effects.)

Upon reopening schools in February 2021 after a prolonged lockdown, Austria started mass rapid antigen testing of all school children twice a week. As symptomatic students could not attend, all students are assumed to have been asymptomatic. Tests were conducted every Monday and Wednesday using the Lepu Medical antigen test, which uses an anterior nasal swab. The first week of school antigen testing resulted in 198 positives among 470,000 tests conducted in Vienna and Lower Austria (in the week February 8 to 12, 2021), yielding a positivity rate of 0.04% [27]. Among these, more than 75% were subsequently confirmed positive using a PCR-test (suggesting a rather high specificity of 99.99%). In weeks two through six, the tests were conducted in all provinces, yielding weekly, Austria-wide positivity rates of 0.04%, 0.065%, 0.09%, 0.08%, 0.08% and 0.08% [28–32]. Across all provinces, the largest increase in positive tests between rounds – on average more than two-fold – occurred between the first and second round of testing (see Supplementary Table 1). This may be due to increased quality of swab taking. In addition, it is possible that COVID-19 incidence is lower in pupils than in the general population, especially right after a lockdown [33,34]. Furthermore, while viral load appears similar between asymptomatic children and adults [35], it is conceivable that school children stay infectious – and antigen-test positive – for a shorter period of time than adults.

The average constant trend among pupils during this period contrasts with the general population, where the effective reproduction number R (based on PCR-incidence) was approximately 1.1 [36]. Therefore, the tested cohort maintained a roughly 10% lower R (note that at the time, testing in the general population was more limited). Rapid antigen tests serve both to detect infectious cases among teachers and pupils, and to identify nascent clusters: in case of a suspected outbreak, a whole class is PCR-tested. While self-administered antigen tests enable the prevention of the majority of future infections [14], there is a small fraction of infectious adults which are not detected using self-administered tests [1,24,37], and this may occur more often for children. A local 'Gurgelstudie' [gargling trial] in March 2021 indicated that the proportion of potentially infectious samples missed by antigen tests may be somewhat higher in children than in adults, although the small sample, coupled with 'relatively high' [sic] $C_t$s does not allow for a robust conclusion (5 out of 14 children with at least one PCR $C_t$ < 30 were also detected by the antigen test) [38]. A detailed study of children and adults seeking testing showed that while antigen-test (BinaxNOW, Abbott) sensitivity was indeed lower for children than for adults (73% vs. 81%), antigen results were positive for all PCR-positive samples from children where viable virus could be isolated [35]. In a school-setting, a large trial in the UK found that bi-weekly antigen testing, followed by daily antigen testing of contacts of identified COVID-19 positive pupils, appears to be similarly effective as isolating all contacts for 10 days, which was the standard policy in the UK at the time [39]. This strongly suggests that properly supervised antigen testing is sensitive enough that a daily test-to-stay policy is a good way to minimize the spread of COVID-19 in schools, while limiting isolation of contacts (same household excluded). Note that in the absence of efficient testing, school closures have been ranked as a very effective measure in reducing spread of COVID-19 [40,41].

While rapid antigen tests were initially recommended mainly to control local outbreaks and to quickly diagnose symptomatic patients [42], use of rapid antigen tests as a general public-health tool has been gaining momentum since the summer of 2020 [3,7,43,44]. The immediate availability and convenience of the rapid antigen tests means that one learns about an active infection promptly, which effectively limits further spread.

Although a recent antigen test (self-administered or not), is a good indicator of infectiousness [24,37,45], it is inevitable that some people will swab incorrectly, leading to false negative results, or fail to quarantine after a positive test, leading to further transmissions. As such, rapid antigen tests must not encourage reckless behavior, but rather enable people to lower their risk of infecting others by testing at least twice per week. Having longer gaps between tests decreases the potential benefit for reducing transmissions [7].

Although vaccination programs are progressing well, vaccine uptake varies greatly across socio-economic groups and is lower in the younger population [46–49]. Vaccination against COVID-19 has only recently been approved and recommended for children; in the fall, schools reopened with pupils largely unvaccinated. Even with a successful vaccination program, vaccine waning presents additional difficulties for achieving full immunity via vaccination [50,51]. We expect that frequent testing will stay exceptionally useful for many months to come: to increase safety in schools and at public events, as well as to help to suppress local outbreaks.

**Ethics approval:**
Ethics committee approval was not required as no new data was collected for this study.

**Acknowledgements:**
Jitka Polechová has received funding from FWF Austrian Science Fund; project P32896-B.

infection. J Infect Dis 2021. https://doi.org/10.1093/infdis/jiab337.

**Supplementary Table 1.**
**Weekly positive rapid antigen tests in Austrian schools, per province.** Approximately 10% of the Austrian population is tested in this program. Primary schools (age 6-10) are open 5 days a week and pupils are tested twice per week. In secondary schools (age 10-15), half of the pupils go to school Monday and Tuesday and are tested on Monday; the other half attend Wednesday and Thursday, and are tested on Wednesday. (Exceptionally, schools use alternate weeks instead). Initially, only Lepu Medical tests were used; lately, some secondary schools started to use Flowflex (ACON) [52]. The frequency of school-testing has now been increased to 3-times a week. All staff wear FFP2 masks, secondary school pupils wear face masks in the classroom, primary school pupils wear masks when indoors outside of their class group. There is some variability to this general guideline for more specialized institutions.

| Week starting | 8.2. | 15.2. | 22.2. | 1.3. | 8.3. | 15.3. | 22.3. |
| --- | --- | --- | --- | --- | --- | --- | --- |
| Wien | 142* | 250 | 258 | 429 | 345 | 606 | 599 |
| Niederösterreich | 56* | 103 | 187 | 209 | 212 | 195 | 234 |
| Oberösterreich | | 43 | 118 | 161 | 217 | 139 | 147 |
| Steiermark | | 43 | 109 | 148 | 122 | 134 | 145 |
| Salzburg | | 19 | 68 | 97 | 90 | 100 | 82 |
| Kärnten | | 40 | 65 | 86 | 84 | 53 | 61 |
| Tirol | | 13 | 42 | 54 | 66 | 43 | 60 |
| Burgenland | | 10 | 34 | 44 | 41 | 49 | 61 |
| Vorarlberg | | 15 | 23 | 19 | 11 | 6 | 16 |
| Total Positive | 198 | 536 | 904 | 1247 | 1188 | 1325 | 1405 |
| Total Tests (appr.) | 470 k | 1.3mil | 1.4mil | 1.4mil | 1.5mil | 1.6mil | 1.7mil |
| **Percent Positive** | **0.042** | **0.041** | **0.065** | **0.089** | **0.079** | **0.083** | **0.083** |

* 75% and 80% of the cases, respectively, were confirmed by PCR in the first week; in later weeks, the % of false positives rose to about 40% [32].